\documentclass[letterpaper]{article}
\usepackage{aaai21}  
\usepackage{times}  
\usepackage{helvet} 
\usepackage{courier}  
\usepackage[hyphens]{url}  
\usepackage{graphicx} 
\usepackage{booktabs}
\usepackage{natbib}  
\usepackage{caption} 
\usepackage{cleveref}
\usepackage{multicol}
\usepackage{subcaption}
\usepackage{todonotes}

\begin{document}

\title{The MeLa BitChute Dataset}
\author {
    Milo Z. Trujillo,\textsuperscript{\rm 1}
    Maur\'{i}cio Gruppi,\textsuperscript{\rm 2}
    Cody Buntain,\textsuperscript{\rm 3} and
    Benjamin D. Horne\textsuperscript{\rm 4} \\ 
}
\affiliations {
    \textsuperscript{\rm 1} Vermont Complex System Center, University of Vermont, Burlington, VT USA\\
    \textsuperscript{\rm 2} Computer Science, Rensselaer Polytechnic Institute, Troy, NY USA \\
    \textsuperscript{\rm 3} College of Information Studies, University of Maryland, College Park, MD USA\\
    \textsuperscript{\rm 4} School of Information Sciences, University of Tennessee Knoxville, Knoxville, TN, USA\\
    milo.trujillo@uvm.edu, gouvem@rpi.edu, cbuntain@umd.edu, bhorne6@utk.edu\\
}


\maketitle

\begin{abstract}
In this paper we present a near-complete dataset of over 3M videos from 61K channels over 2.5 years (June 2019 to December 2021) from the social video hosting platform BitChute, a commonly used alternative to YouTube. Additionally, we include a variety of video-level metadata, including comments, channel descriptions, and views for each video. The \texttt{MeLa-BitChute} dataset can be found at: \url{https://dataverse.harvard.edu/dataset.xhtml?persistentId=doi:10.7910/DVN/KRD1VS}.
\end{abstract}


\section{Introduction}
\citet{Wilson2021} define ``alt-tech'' spaces as alternative, non-mainstream platforms that exist largely in reaction to perceived risks of censorship in mainstream spaces. These alt-tech platforms have received significant attention from researchers, practitioners, and even policy makers, due to their role in producing, spreading, and conserving anti-social content. This anti-social content ranges from political disinformation to health-related conspiracy theories to violent hate speech. 

To this end, researchers have striven to study alt-tech platforms and to build large datasets around those platforms; e.g., Gab \cite{zannettou2018gab}, Gettr \cite{Paudel2021}, Parler \cite{Aliapoulios2021}, Dissenter \cite{Rye2020}, and Telegram \cite{Junior2021}. However, one platform in this ecosystem that has, for the most part, lacked study is BitChute, an alternative to YouTube. Studies on BitChute are rare because data collected from the platform is rare, as BitChute does not have a publicly available API like other social media platforms. This limitation makes data collection a significant hurdle for researchers.

Despite this difficulty, the platform is deserving of study. Just as other major alt-tech platforms, BitChute plays a critical role in harboring anti-social content and communities \cite{trujillo2020bitchute}. Most famously, BitChute was a safe haven for the viral, COVID-19 conspiracy theory film, \emph{Plandemic}, which was quickly removed from Facebook, YouTube, and Twitter \cite{Kearney2020, Buntain2021}. As shown in \citet{Rogers2020}, these alternative spaces do not exist in isolation; rather, users of and audiences in alt-tech spaces operate across several of these platforms simultaneously. Hence, these spaces need to be studied holistically to understand how extremists and far-right individuals leverage the many affordances available across these alt-tech platforms. This need is apparent, as \citet{Doesburg2021} demonstrates that BitChute is already one of the most popular alt-tech domains shared in Telegram, and \citet{Rogers2020} shows that BitChute is one of the main destinations for internet celebrities who have been deplatformed on the mainstream platforms. As the broader information ecosystem cannot be fully understood without YouTube, so too must one understand BitChute's role to understand the alt-tech ecosystem.  

In this paper, we present the \texttt{MeLa-BitChute} dataset to help fill this gap. The dataset contains data from 3,036,190 videos, 61,229 channels, and 11,434,571 comments between June 28th, 2019 and December 31st, 2021. This dataset provides timestamped activities and estimates on views for the majority of channels and videos on the platform, allowing researchers to align BitChute videos with behavior on other platforms. Therefore, this dataset can facilitate both studies of BitChute in isolation and studies of BitChute's role in the larger ecosystem. 

In the remainder of this paper, we describe the data collection methodology behind \texttt{MeLa-BitChute}, publicly-available data formats and documentation, evaluations of the dataset's completeness, and an extensive discussion on use cases. While we do discuss some characteristics of BitChute in this paper, we recommend reading \citet{trujillo2020bitchute} for a more complete examination of the platform's history and early content.

\begin{figure*}[h]
    \centering
    \includegraphics[width=14cm]{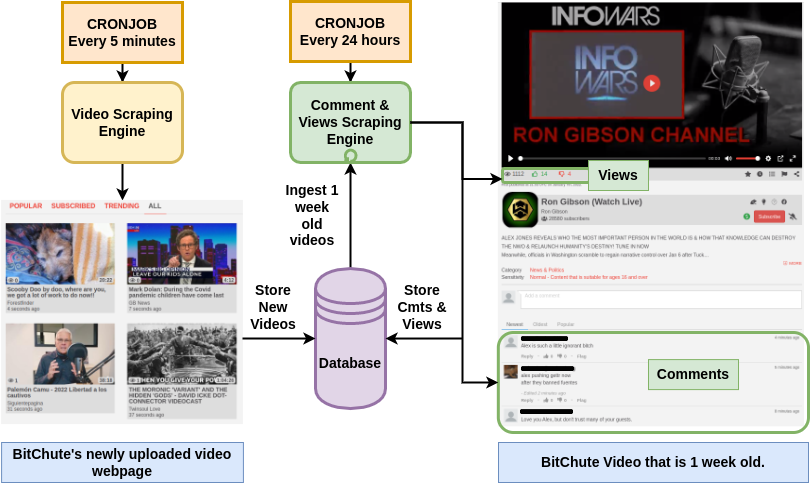}
        \caption{The core data collection is done in two parallel parts: 1. Every 5 minutes we collect newly uploaded videos from the BitChute video stack. 2. Every 24, hours we loop through all videos that are at least one week old, go to the video page, and scrape views- and comments data. The third component, not depicted here, is the collection of channel description, which is scraped once for every new channel in the database.}
    \label{fig:flow}
\end{figure*}

\begin{figure}[h]
    \centering
    \includegraphics[width=8cm]{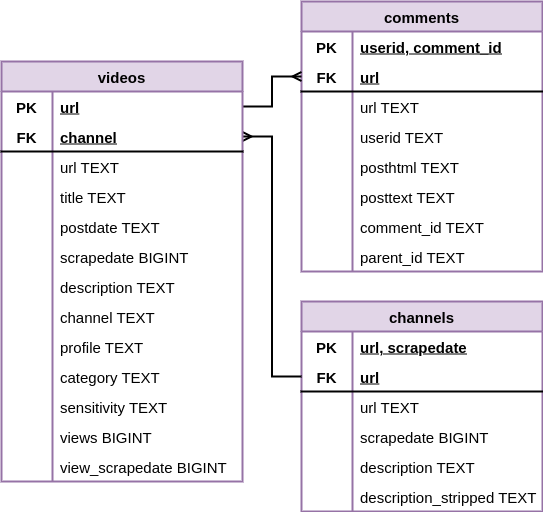}
        \caption{Database schema for SQLite3 format. The video URLs are used as the primary key for the videos table, while a composite key of the commenter user ID and the individual comment ID are used as the primary key of the comments table, and a composite key of the channel URL and the timestamp the channel description was scraped are used as the primary key of the channels table.}
    \label{fig:db}
\end{figure}

\section{Data Collection and Infrastructure}
BitChute does not have a published API. Therefore, we build a custom collection engine that uses web-scraping and parsing at regular intervals. Broadly, using this infrastructure, we collect data on four entities within BitChute: videos, video views, comments, and channels. Figure \ref{fig:flow} shows the high-level flow chart of this collection infrastructure.

\subsection{Videos} 
\begin{figure*}[h]
    \centering
    \includegraphics[width=14cm]{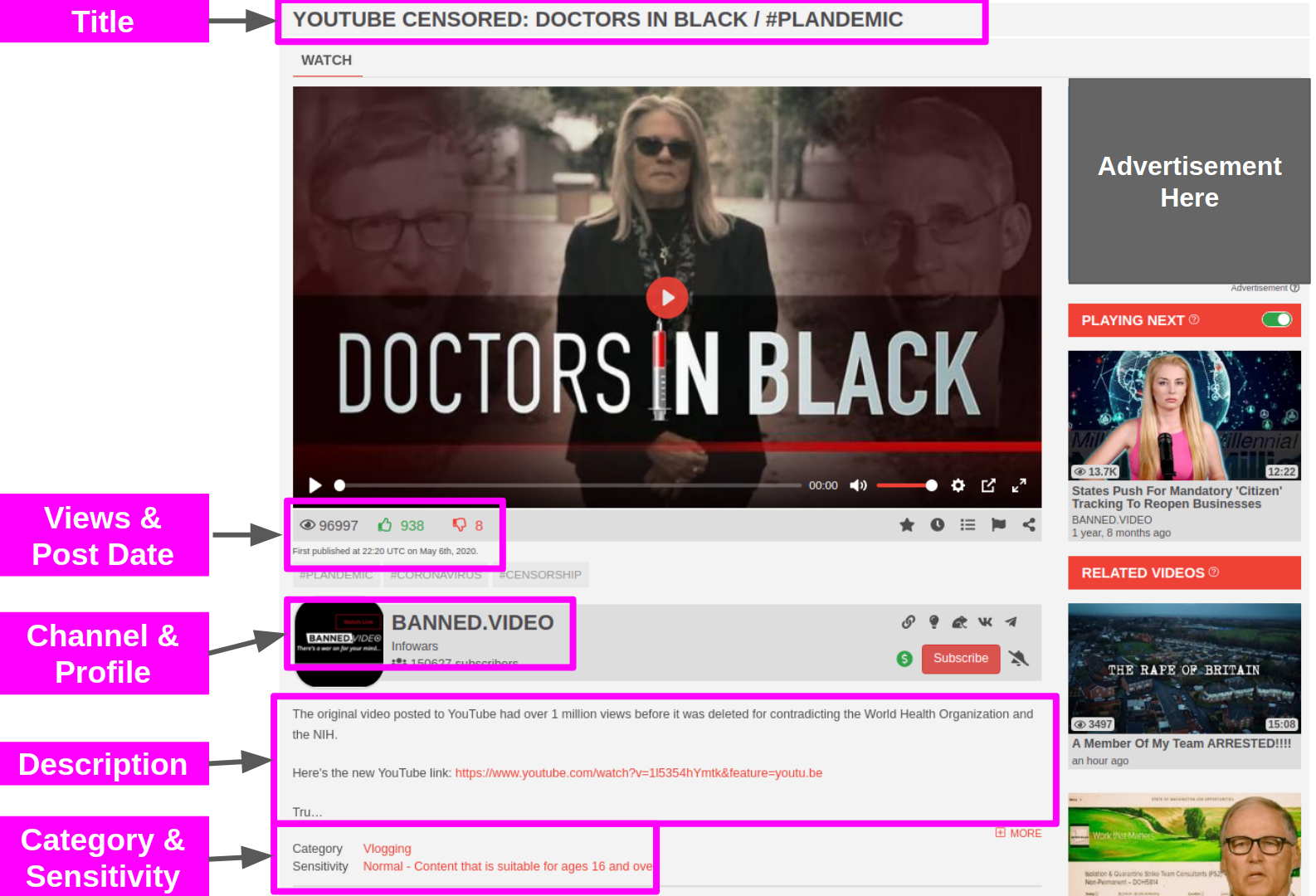}
        \caption{Example of metadata collected for each video. Note, the channel of the video is ``BANNED.VIDEO'' while the profile is ``Infowars.'' The Infowars profile has several channels, including an Infowars channel and ``The Alex Jones Show.''}
    \label{fig:video_example}
\end{figure*}

The core part of the data collection is video metadata. To collect video data, we utilize a web-page on BitChute that displays newly uploaded videos in a stack (see Figure \ref{fig:flow} for an example of this page). This page is scraped every five minutes, recording the URLs of videos we have not yet seen. For each newly-uploaded video, we visit the video URL, parsing the web interface to extract the video title, description, uploader, channel, video category, sensitivity rating, and exact upload date. We record these results in a PostgreSQL database. An example of the video interface that data is scraped from can be found in Figure \ref{fig:video_example}.

\subsection{Comments and Views}
Since we collect videos within 5 minutes of being uploaded to the platform, we must wait to collect audience engagement data. To this end, we have a second, concurrent process, which queries the database for a list of videos uploaded at least one week ago that have not yet been re-examined. This script then visits each video URL again, this time recording the number of views and scraping all of the comments that the video has received. If the video is no longer available, we record whether it has been retracted by the uploader or removed by BitChute moderation.

When scraping the comments, we collect the comment author, comment text, creation date, and scrape date. 


\subsection{Channels}
Lastly, to collect description data on the channels, we have a third concurrent process that queries the database for a list of channels that have yet to have their description information collected. This script then visits each channel `About' page to scrape the HTML for an author-provided description, if available. 

\subsection{Collecting Dynamically Loaded Data} \label{sec:dynamic}
While most of the metadata we are interested in is available through HTML parsing, the view counts and comments are loaded dynamically. We initially overcame this by automating a web browser with Selenium, so that BitChute's JavaScript could run without modification. Using this method, we could extract the view count and comments from the browser's DOM. However, automating a browser scaled poorly, so we instead determined what HTTP requests the site JavaScript was making, and automated those requests ourselves to retrieve views and comments.

\subsubsection{Collecting Dynamically Loaded Views}
First, to retrieve views, we visit the video URL and record a CSRF token\footnote{Cross-Site Request Forgery Token, a type of cookie typically used to prevent attackers from tricking a browser into making malicious HTTP requests}. Then when visiting \texttt{/\{video-URL\}/counts/}, we present the CSRF token and collect the loaded views data.

\subsubsection{Collecting Dynamically Loaded Comments}
Second, to retrieve comments, we must rely on BitChute's comment infrastructure, which changed throughout our collection timeline. Originally, from 2017 to around September 2020, BitChute contracted the third-party blog comment hosting service, Disqus, to provide commenting infrastructure for their website. After Disqus terminated the contract, BitChute implemented their own commenting software, ``CommentFreely." CommentFreely is open source, and can be examined at \url{https://github.com/BitChute/commentfreely}.


When Disqus was being used, we derived the Disqus URL from the BitChute URL, then visited the Disqus page, which included comments as both HTML and JSON for easy parsing. Once BitChute switched to their own CommentFreely infrastructure, we retrieved comments as follows: 1. Parse the CommentFreely JavaScript embedded in the video page to find a unique ``video token." 2. Make a POST request to \url{https://commentfreely.bitchute.com/api/get_comments/} containing the video token. 3. Parse the response JSON.

\section{Publicly Available Data Formats}  
In order to accommodate the largest audience possible, we provide two widely-used data formats. 

\subsection{SQLite3 Database}
The first format is an SQLite3 database with three tables: videos, comments, and channels. The schema for this database can be found in Figure \ref{fig:db}. While our collection engine stores data in a PostgreSQL database, we convert it to an SQLite3 database to allow researchers to use the data without any database server setup. 

The primary table in the database is the \texttt{videos} table, which includes the video URL, title, date the video was posted, timestamp when video was scraped, a description of the video if provided by the uploader, the profile of the user who posted the video, the channel that posted the video (see Appendix Table \ref{tbl:vids} for description of profile vs. channel), the user picked topical category for the video, the user picked sensitivity for the video, the number of times the video is viewed, and a timestamp of when the views were scraped. 

The second table is the \texttt{comments} table, which contains the video URL that the comment is under, the user ID of the commenter, the comment ID for the individual comment, the parent comment ID if it is a nested comment, the comment HTML, and the comment text. 

The third table is the \texttt{channels} table, which contains the URL to the channel, a the HTML of the channel description if the channel owner provides one, the text of the channel description if the channel owner provides one, and a timestamp of when the channel data was scraped.

A detailed description of each data column can be found in Appendix \ref{appendix:coumns}.

\subsection{CSV}\label{sec:csv}
The second format in which we provide the dataset is a set of Comma-Separated Value (CSV) files. We provide three CSV files, one for each table in the database: videos, comments, and channels. The columns in each CSV file are the same as the columns in each corresponding SQLite3 database table. 

\subsection{FAIR Principles}
We are careful to ensure that the \texttt{MeLa-BitChute} dataset follows FAIR principles\footnote{\url{https://www.force11.org/group/fairgroup/fairprinciples}}. 
\begin{itemize}
    \item Findable - The dataset is persistently stored on Harvard Dataverse, is documented, and described with rich metadata. 
    \item Accessible - The dataset is freely and publicly accessible through Harvard Dataverse's GUI, is stored in two widely-used formats, and comes with an example Python script for data extraction and use.
    \item Interoperable - The dataset can be parsed automatically using standard languages (Python, SQL, R), and can be parsed by human annotators using the CSV formats provided. 
    \item Re-usable - The data can be reused for many types of studies, given the breadth of the collection. The data can be paired and augmented with other social media datasets for rich studies of alt-tech, deplatforming, disinformation spread and more. See Section \ref{sec:use} for an in-depth discussion of these use cases. Given that the URLs are stored and rich metadata is well-documented, provenance is maintained.
\end{itemize}

\subsection{Ethical Considerations of Stored Data}
The key ethical consideration with the collection and sharing of this dataset is the privacy of the platform's producers and consumers. While we considered anonymizing the content producers (channels/profiles), we choose not to as their identity is important to understand both activity within the platform and studies across other platforms. Censoring the channel and video URLs would reduce the dataset's usefulness in the research community and would destroy the provenance of the data. Furthermore, since the content producers on BitChute post videos for public consumption, they should not have the expectation of anonymity. 

On the other hand, we do choose to anonymize the comment posters. We have assigned each commenter a unique ID, by creating a salted hash of their account information. This allows researchers to identify all comments made by a particular author, without revealing the username of the author. We do this because commenters have a greater expectation of privacy than content publishers, and the right to be forgotten. Furthermore, the importance of tracking influential content creators across multiple platforms does not apply to commenters. 



\begin{figure*}[h]
    \centering
    \begin{subfigure}{.33\textwidth}
        \centering
        \includegraphics[width=5.2cm]{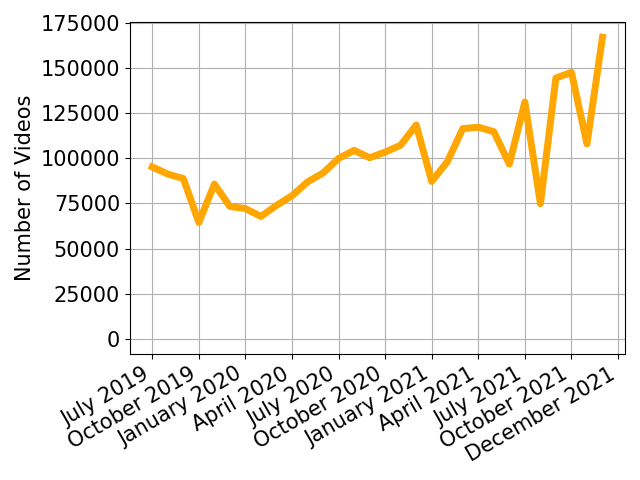}
        \caption{Videos published per month}
        \label{fig:vids_per_month}
    \end{subfigure}
    \begin{subfigure}{.33\textwidth}
        \centering
        \includegraphics[width=5.2cm]{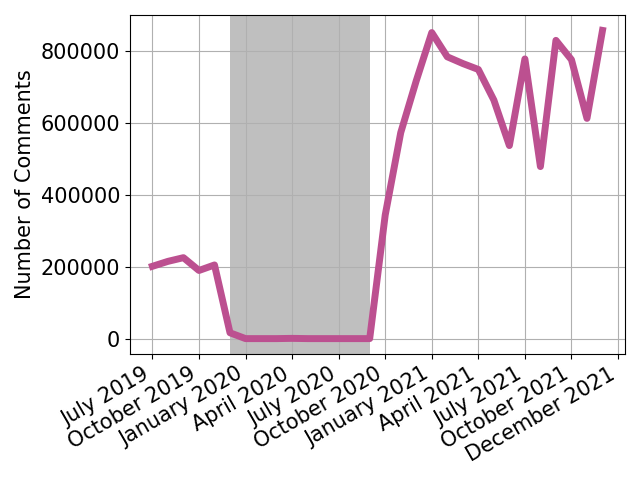}
        \caption{Comments posted per month}
        \label{fig:cmts_per_month}
    \end{subfigure}
    \begin{subfigure}{.33\textwidth}
        \centering
        \includegraphics[width=5.2cm]{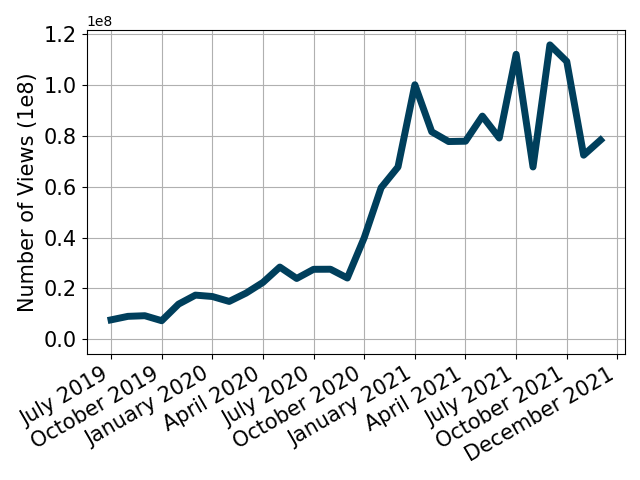}
        \caption{Views per month}
        \label{fig:views_per_month}
    \end{subfigure}
\caption{In \textbf{(a)} we show the number of videos published each month. In \textbf{(b)} we show the number of comments posted in each month. Note, the grey background indicates dates where we were not able to collect comments. Please see discussion in Section \ref{sec:dataquality}. In \textbf{(c)} we show the number of video views per month. Note both the growth in videos and views collected per month (a and c) and the growth in Google Trends interest in BitChute in Figure \ref{fig:rends}.}
\end{figure*}

\begin{figure}[h]
    \centering
    \includegraphics[width=7cm]{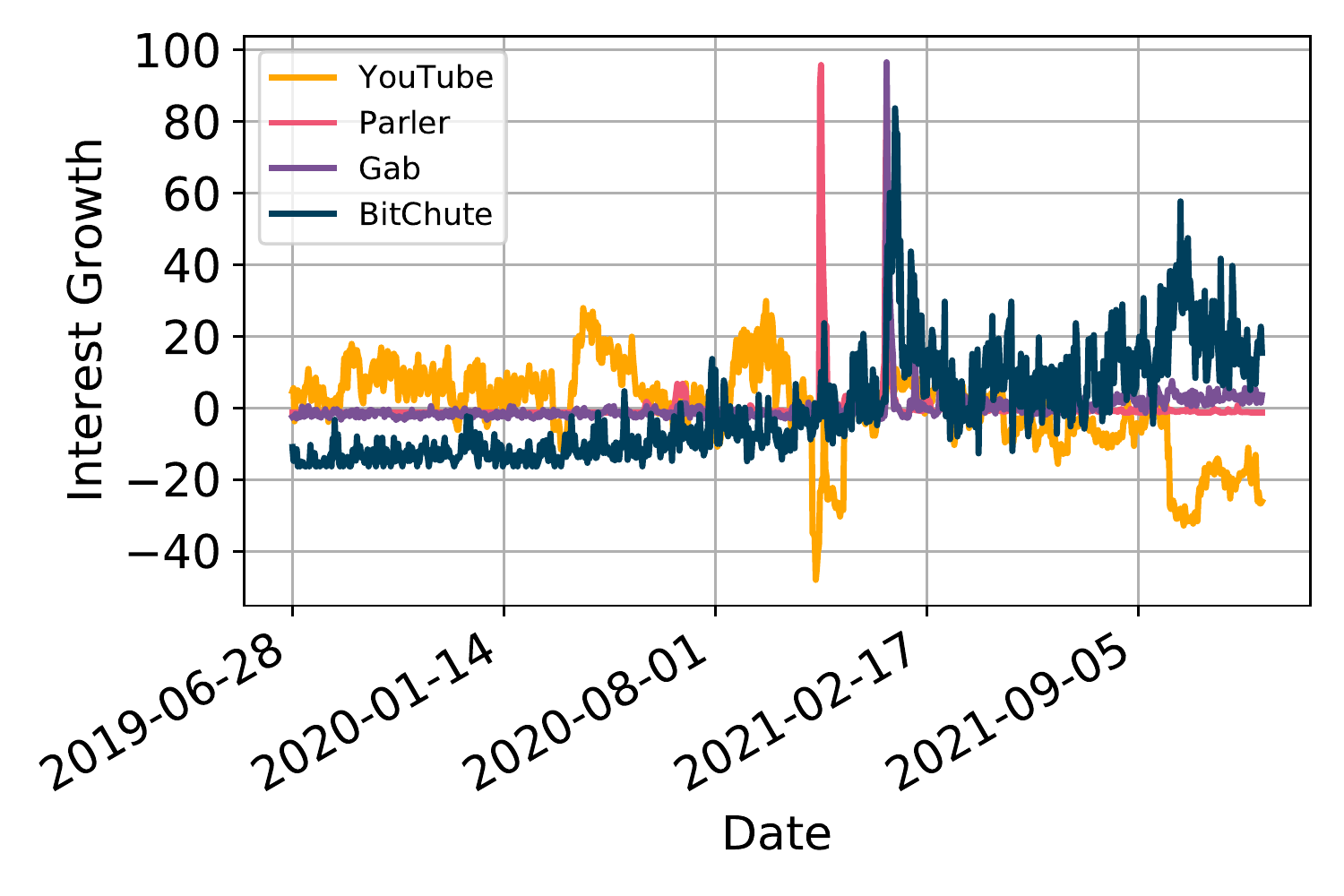}
        \caption{Google Trends interest across the dataset's timespan, where the mean is shifted to 0 for comparison.}
    \label{fig:rends}
\end{figure}


\begin{figure*}[h]
    \centering
    \begin{subfigure}{.45\textwidth}
        \centering
        \includegraphics[width=7cm]{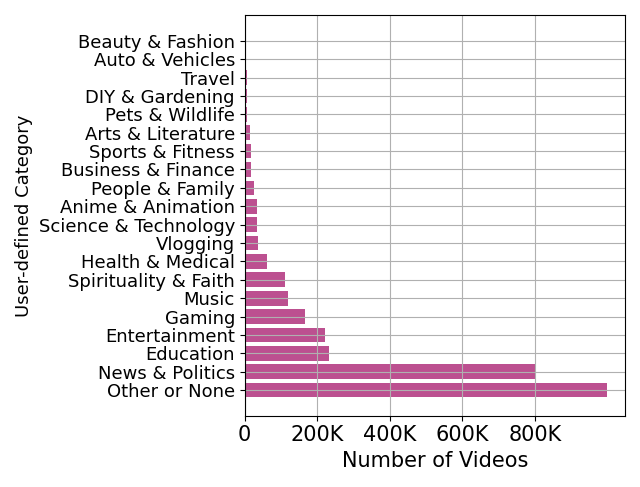}
        \caption{Top 20 user-defined categories by number of videos}
        \label{fig:top20cats}
    \end{subfigure}
    \begin{subfigure}{.45\textwidth}
        \centering
        \includegraphics[width=7cm]{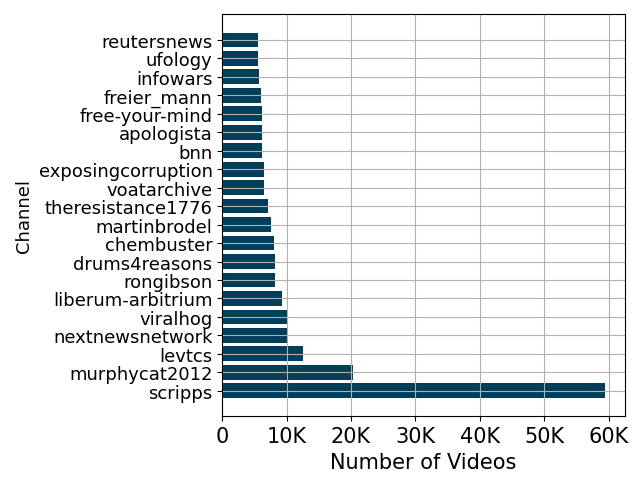}
        \caption{Top 20 channels by number of videos}
        \label{fig:top20chans}
    \end{subfigure}\\
    \begin{subfigure}{.45\textwidth}
        \centering
        \includegraphics[width=7cm]{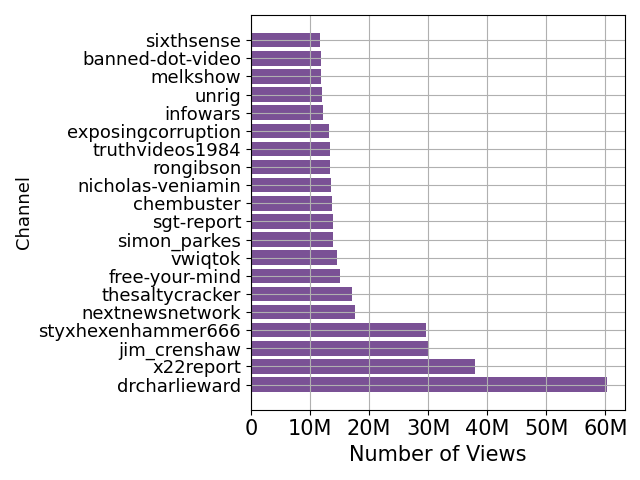}
        \caption{Top 20 channels by number of views}
        \label{fig:top20chans_views}
    \end{subfigure}
    \begin{subfigure}{.45\textwidth}
        \centering
        \includegraphics[width=7cm]{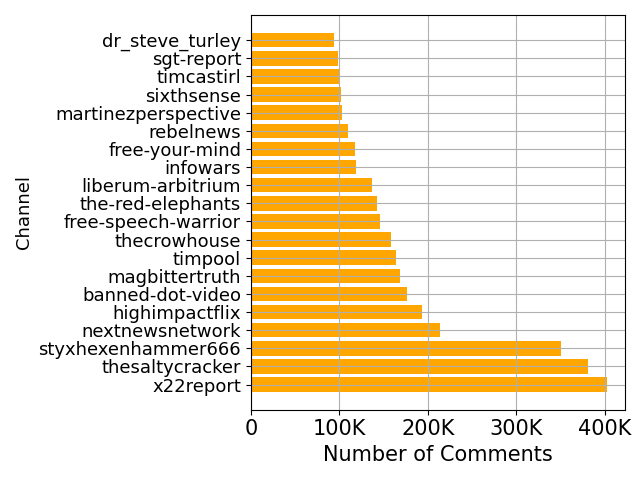}
        \caption{Top 20 channels by number of comments}
        \label{fig:top20chans_cmts}
    \end{subfigure}
\caption{In \textbf{(a)} we show the top 20 user-picked video categories by number of videos. Note, just as described in \cite{trujillo2020bitchute}, `Other' was originally the default category for all uploaded videos. However, in September 2020 BitChute changed the `Other' category to `None'. We show the combined number of videos across those two categories. In \textbf{(b)} we show the top 20 channels by number of videos, in \textbf{(c)} by number of views, and in \textbf{(d)} by number of comments.}
\end{figure*}

\section{Evaluation of Data Completeness}\label{sec:dataquality}
Given the complexity of collecting this data, it is important to clearly document where we are confident in the data completeness and where we are not. To this end, below we discuss several known data outages and caveats.

\subsubsection{Known collect server outages}
During the collection of this data, we documented several collection outages due to either BitChute itself or due to issues at our server location, which changed several time during the timeline. Below is a list of documented shutdowns of our data collection:

\begin{itemize}
    \item October 11th, 2019 - Our collection server's IP address was blocked from accessing BitChute.
    \item October 23rd, 2019 and November 18th, 2019 - Our collection server lost connectivity during the Pacific Gas \& Electric preemptive power shutdowns.\footnote{\url{www.latimes.com/california/story/2019-11-18/another-power-outage-pge-may-shut-down-grid-northern-california}} Our sever was located in California at the time.
    \item November 22nd, 2019 - BitChute itself was down. The cause is unknown.
    \item November 3rd, 2020 - BitChute itself was down due to one of their service providers cutting their account off.
    \item May 22nd, 2021 to May 24th, 2021 - A power outage in the block of the University of Tennessee Knoxville's campus where our server was located. The cause of the outage was a mouse entering switch gear at the iconic American football stadium, Neyland Stadium.
    \item June 12th, 2021 to June 14th, 2021 - Another power outage in the block of the University of Tennessee Knoxville's campus where our server was located.
\end{itemize}

In most of these cases, we were able to recover videos published during the outages, but we cannot guarantee we recovered them all. Given how short these outages were, we are confident that the number of videos collected over-time is near-complete. 

\subsubsection{Known missing comment data}
As discussed in Section \ref{sec:dynamic}, BitChute's comment infrastructure changed throughout our collection timeline. Since our collection method is highly dependent on the structure of BitChute's platform, these changes directly impacted our ability to collect comments. Originally, BitChute used third-party Disqus, then created their own called CommentFreely after Disqus terminated its contract with BitChute. During this transition, multiple structural changes by BitChute made it impossible to collect comments. Furthermore, we were unable to recover comment data because the comments made during the Disqus era were removed by BitChute and no longer exist. In Figure \ref{fig:cmts_per_month}, dates where comment data is missing due to this are shown in grey. Based on the average number of comments per video, we estimate that we missed at maximum 8.8M comments. This estimate is likely an over-estimate given the platform growth during this time-span. In total, 22.9\% (697K) of the videos collected were published during the time-span where comments could not be collected. Therefore, we are confident in conclusions being drawn from the temporal patterns of the videos throughout the collection timeframe but not in conclusions drawn from the temporal patterns of  comments during 2020.

\subsubsection{Caveats on views and comments data}
Again, due to BitChute's multiple commenting systems, and each system having a different comment ID format. Comment IDs should be assumed to be text, not numeric, and may be NULL. Author IDs cannot link authors from before and after the Disqus to CommentFreely change.

Unfortunately, due to the software changes on BitChute, some of the views and comments are gathered later than one week after video publication. Since our comment data includes the creation date of each comment, researchers can choose to filter out comments more than a week newer than their corresponding videos to obtain a consistent ``one week later" view of the dataset.






\section{Dataset Use Cases}\label{sec:use}

As one of the main video-sharing platforms in the growing alt-tech ecosystem, researchers are increasingly interested in BitChute and its users as lenses for studying these alt-tech spaces.
This interest is evident in Google Scholar data, where the number of articles mentioning BitChute has doubled annually over the past two years.
Despite this interest, limited data resources exist to support these studies.
Given the limited availability of such data, the \texttt{MeLa-BitChute} dataset can facilitate such studies in a multitude of ways, which we outline below.

\paragraph{Content Moderation and Deplatforming}

As social media spaces have become central sources of information despite the prevalence of misinformation and efforts to manipulate audiences \cite{Tucker2018}, the moderation tools and interventions used to ensure the safety of online audiences have likewise become critical aspects of the information ecosystem.
``Deplatforming'', or removing/suppressing content or individuals from an online platform, has emerged as a popular moderation intervention across both the mainstream social media platforms (e.g., Twitter, Facebook, and Reddit) \cite{MastersofMedia2020,VanDijck2021,Rogers2020} and infrastructure platforms (e.g., Amazon Web Services and Google/Apple app stores) \cite{Buckley2021}.
Assessing impact of these interventions can no longer be done within the context of a single platform, however, as the proliferation of alt-tech platforms has provided new spaces for creators to circumvent moderation.
Of particular concern is whether deplatforming pushes individuals from mainstream spaces to these more extreme spaces, thereby increasing exposure to toxic and extreme content in spaces like BitChute \cite{Buntain2021b}.

Researchers have already begun investigations into the cross-platform effects of deplatforming, especially between YouTube and BitChute: 
\citet{Buntain2021} examines YouTube's de-recommendation strategy, finding that removing recommendations to misinforming content appears to suppress its sharing on Twitter and Reddit, but little is known about whether this change simply moves that de-recommended content to alt-tech spaces like BitChute.
Similarly, in the aftermath of the ``Great Deplatforming'' around the January 6 attack on the US Capitol \cite{Bond2022}, \citet{Buntain2021b} shows interest in alt-tech platforms like Gab and BitChute has increased.
\citet{Rauchfleisch2021}, on the other hand, show that, only about 20\% of far-right channels on YouTube that were deplatformed between 2018-2019 had a BitChute presence at the end of 2019.
The ecosystem has evolved since these studies, however, and the \texttt{MeLa-BitChute} dataset enables these and new studies of whether and how BitChute creators and audiences respond to these deplatforming efforts. 
Such questions include whether creators push new content to their BitChute channels or whether BitChute audiences grow and videos are seen or shared more in the aftermath of deplatforming -- as suggested with the Plandemic video \cite{Bellemare2020,Buntain2021}.

As a simple example of the platform's evolution since these studies, one can compare popular channels in Figures \ref{fig:top20chans_views} and \ref{fig:top20chans_cmts} (across 2019-2021) to those in 2019, as documented in \citet{trujillo2020bitchute}.
We see some overlap in the channels with high engagement from late-2019 (\textit{infowars}, \textit{rongibson}, \textit{x22report}, \textit{styxhexenhammer666}, \textit{nextnewsnetwork}), but new channels have emerged as highly engaged in this dataset, such as \textit{drcharlieward}, \textit{chembuster}, and \textit{sixthsense}. Many of these new top channels are focused on health and vaccination conspiracy theories, where as the top channels in late-2019 were mostly focused on far-right politics.

\paragraph{Conspiracy and Political Misinformation}

Like other alt-tech spaces, BitChute has a high proportion of political content, much of which focuses on conspiracy theories and otherwise politically extreme \cite{trujillo2020bitchute,David2020}.
These alt-tech spaces are especially popular among far-right audiences, who often use these platforms to share and amplify right-wing conspiracies and misinformation \cite{Freelon2020a}.
For example, despite moderation action by the mainstream platforms to suppress election-related misinformation in the aftermath of the January 6 attack, such conspiracies have flourished in BitChute \cite{Heilweil2021}.
The QAnon constellation of conspiracy theories is similarly popular on BitChute \cite{trujillo2020bitchute}, as early research is already examining QAnon's use of the platform for sharing its increasingly popular messages  \cite{Forberg2021,Hoseini2021}.
The \texttt{MeLa-BitChute} dataset is valuable in this space, as much of the content contained therein would be removed or suppressed on other platforms, and the alt-tech spaces available are primarily text-oriented.
Consequently, the content in BitChute and captured by the \texttt{MeLa-BitChute} dataset represents a unique multi-modal resource that can provide insight into current and emerging topics of conspiracy and political misinformation.


\paragraph{Supply, Demand, and Health Misinformation}

Following the release of the misinformation-laden \emph{Plandemic} film on social media, Facebook, YouTube, and Twitter responded quickly to limit its spread \cite{Kearney2020}.
The film contained many unfounded conspiracy theories about the origin of the COVID-19 pandemic, leading these mainstream platforms to ban or suppress it as it violated policies on misinformation related to public health.
Despite these interventions, the film remained widely available on alt-tech spaces and on BitChute in particular \cite{Bellemare2020}, and analysis in \citet{Buntain2021} suggests BitChute received more traffic in response, as interest in the film drove viewers to platforms that were willing to host it.
Viewed through the supply-and-demand framework in \citet{Munger_2019}, as the mainstream platforms limit  COVID-19 misinformation in their spaces, the supply of this content moves to these alt-tech spaces.
If demand remains constant, interventions by the mainstream platforms may then push audiences toward these alternative spaces, where both misinformation and extreme rhetoric are more common.
Two questions then emerge: First, is health misinformation becoming more popular in BitChute and related spaces as Facebook, YouTube, and others increasingly suppress that content locally, and second, how are content producers on BitChute, Gab, etc. responding to this influx of demand -- e.g., are they producing more such content?
The \texttt{MeLa-BitChute} dataset provides insight into these questions, first by allowing researchers to evaluate trends in engagement and viewership, and second through longitudinal data of content creators and the videos they post to their channels over time.

\paragraph{Alternative Monetization}

Related to the supply and demand questions above, creators have multiple ways to monetize their supply of content.
YouTube's Partner Program, for example, pays creators a portion of advertising revenue based on views \cite{Kopf2020}. 
When YouTube deplatforms or otherwise suppresses content, however, creators are ``demonetized'' and lose out on this revenue stream.
Despite these interventions, creators have found ways to circumvent deplatforming by posting their more violative content on spaces like BitChute and sharing trailers to this content on mainstream platforms \cite{trujillo2020bitchute}.
To this end, an ecology of alternative monetization schemes now exists, allowing creators to monetize their content through other means, such as donation, cryptocurrency, affiliate marketing, or merchandise \cite{277236}.
BitChute also supports a variety of these monetization options, both through on-platform advertising and integration with donation-based platforms (e.g., Patreon, PayPal, and others).
\citet{Warreth2021} details how the far-right and extremist groups use these alternatives, especially cryptocurrency as funding sources.
Via the the \texttt{MeLa-BitChute} dataset, researchers can study how BitChute's alternative content is monetized through these alternative means.

\paragraph{Hate and Online Extremism}

A rich body of work has examined hateful and radicalizing content in the mainstream platforms.
Evidence shows such content on Facebook has contributed to violence and radicalization; e.g., Jihadist groups have used the platform to radicalize potential recruits \cite{10.2307/26463917}, anti-refugee sentiment predicts criminal acts targeting refugees in otherwise similar communities \cite{Muller2021}, and anti-Muslim sentiment has been used to stimulate fear and violence against Muslim communities \cite{10.2307/26508117}.
Alt-tech spaces like Gab are known to have high proportions of hate speech \cite{zannettou2018gab}, and BitChute is no exception, with much of its content containing hateful and extreme rhetoric, often antisemitic or racist in nature \cite{trujillo2020bitchute,David2020,Papadamou2021}.
Hate speech need not be confined to textual modalities either, as the Anti-Defamation League has shown through its database of hateful symbols \cite{AntiDefamationLeague2022}, and the data contained in the \texttt{MeLa-BitChute} dataset may yield data for studying hateful imagery, as the platform's core affordance is video sharing.
Prior work has also shown such propensity towards hate is both indicative and predictive of violent acts \cite{Abdalla2021}.
Coupling these works with the \texttt{MeLa-BitChute} dataset can provide a lens through which this hateful content can be studied and assessed for potential harm or as an indicator for new violent attacks.
Likewise, the multi-year timeframe covered by the \texttt{MeLa-BitChute} dataset may allow researchers insight into radicalization processes among BitChute content creators and their audiences.
Research enabled by this dataset those studying terrorism, who have expressed a desire for more data-driven analyses of hateful and extreme online content, actors, and audiences \cite{Pelzer2018}.


\section{Related Datasets}


Many available social media datasets are topically focused -- e.g., covering disasters \cite{Olteanu2014,Buntain2021a}, COVID-19 \cite{10.1145/3404820.3404823}, and hate/harassment \cite{ICWSM1715665} -- with new datasets regularly released from spaces like SemEval\footnote{\url{https://semeval.github.io/}} or the annual Text Retrieval Conference (TREC).
While these datasets are valuable resources for understanding specific phenomena, they provide limited general insight into trends across the full platforms.
To study new questions not covered by these topic-specific datasets, researchers often need to build new datasets, which introduces confounders when these datasets need to be collected retrospectively.
Such problems include limits on search timeframes (i.e., one can only go back so far), memory-hole problems \cite{Marshall2020} (i.e., relevant information may have been deleted from the target platform, especially a risk for anti-social behavior), or API changes.
The \texttt{MeLa-BitChute} dataset outlined herein solves these issues by providing a reusable, general collection that characterizes the full BitChute platform over a multi-year period, facilitating research questions in a consistent context.

These issues can be addressed with sufficiently large samples of the target platform, and it is this type of sample that the \texttt{MeLa-BitChute} dataset provides.
Similar platform-wide datasets are available for other platforms, most notably the Pushshift.io Reddit dataset  \cite{Baumgartner_Zannettou_Keegan_Squire_Blackburn_2020}.
Unlike Reddit, BitChute is part of the alt-tech space and fits into a constellation of recent work studying these alternative spaces, including releases from Parler \cite{Aliapoulios2021}, Gab \cite{Fair2019}, and Mastodon \cite{Zignani2019}.
Paralleling YouTube's centrality in the mainstream information ecosystem, BitChute and its video-hosting is likewise a core element of the alt-tech space and is often a highly shared domain in other fringe platforms and Telegram channels.
Taken together, these collections provide a crucial cross-platform view into the ecosystem.

%

Separate from the above published datasets, ``hacktivists'' have released several large-scale datasets from alt-tech spaces, including Gab and Parler \cite{Sharma2021}.
These datasets, such as those hosted on \url{DDoSecrets.com} and Wikileaks, provide more insight into these platforms but at significant ethical and intellectual risk.
In particular, these leaked datasets often contain private data, such as direct messages, that users would not intend for public distribution.
The \texttt{MeLa-BitChute} dataset instead exclusively contains public-facing data, which may miss out on important activity like collusion, brigading, radicalization, or other anti-social behaviors.
These two sources come with different insights and ethical considerations, and we leave questions about which source is most appropriate to future researchers.

\section{Conclusion}
In this paper, we presented a dataset covering 3M+ videos, 61K+ channels, and 11.4M+ comments from the alt-tech, social, video-hosting platform BitChute. We provided an in-depth description of our custom built data collection infrastructure, documentation on the stored data, and a discussion of potential use cases for the dataset. We argued due to the difficulty of data collection, the academic literature is lacking diverse, large-scale studies of BitChute and its role in the alt-tech ecosystem. By filling this gap, researchers can gain a holistic-view of the alt-tech environment and the potential public harms fueled by BitChute. The \texttt{MeLa-BitChute} dataset and sample code can be found at: \url{https://dataverse.harvard.edu/dataset.xhtml?persistentId=doi:10.7910/DVN/KRD1VS}.

\begin{small}
\bibliography{scibib}

\begin{thebibliography}{44}
\providecommand{\natexlab}[1]{#1}
\providecommand{\url}[1]{\texttt{#1}}
\providecommand{\urlprefix}{URL }
\expandafter\ifx\csname urlstyle\endcsname\relax
  \providecommand{\doi}[1]{doi:\discretionary{}{}{}#1}\else
  \providecommand{\doi}{doi:\discretionary{}{}{}\begingroup
  \urlstyle{rm}\Url}\fi

\bibitem[{Abdalla, Ally, and Jabri-Markwell(2021)}]{Abdalla2021}
Abdalla, M.; Ally, M.; and Jabri-Markwell, R. 2021.
\newblock {Dehumanisation of ‘Outgroups' on Facebook and Twitter: towards a
  framework for assessing online hate organisations and actors}.
\newblock \emph{SN Social Sciences} 1(9): 1--36.
\newblock \doi{10.1007/s43545-021-00240-4}.

\bibitem[{Aliapoulios et~al.(2021)Aliapoulios, Bevensee, Blackburn, Bradlyn,
  Cristofaro, Stringhini, and Zannettou}]{Aliapoulios2021}
Aliapoulios, M.; Bevensee, E.; Blackburn, J.; Bradlyn, B.; Cristofaro, E.~D.;
  Stringhini, G.; and Zannettou, S. 2021.
\newblock {A Large Open Dataset from the Parler Social Network}.
\newblock \emph{Proceedings of the International AAAI Conference on Web and
  Social Media} 15(1): 943--951.

\bibitem[{{Anti-Defamation League}(2022)}]{AntiDefamationLeague2022}
{Anti-Defamation League}. 2022.
\newblock {Hate on Display: Hate Symbols Database}.
\newblock Technical report, AntisDefamation League, New York.
\newblock \urlprefix\url{https://www.adl.org/hate-symbols}.

\bibitem[{Baumgartner et~al.(2020)Baumgartner, Zannettou, Keegan, Squire, and
  Blackburn}]{Baumgartner_Zannettou_Keegan_Squire_Blackburn_2020}
Baumgartner, J.; Zannettou, S.; Keegan, B.; Squire, M.; and Blackburn, J. 2020.
\newblock The Pushshift Reddit Dataset.
\newblock \emph{Proceedings of the International AAAI Conference on Web and
  Social Media} 14(1): 830--839.
\newblock
  \urlprefix\url{https://ojs.aaai.org/index.php/ICWSM/article/view/7347}.

\bibitem[{Bellemare, Nicholson, and Ho(2020)}]{Bellemare2020}
Bellemare, A.; Nicholson, K.; and Ho, J. 2020.
\newblock {How a debunked COVID-19 video kept spreading after Facebook and
  YouTube took it down}.
\newblock
  \urlprefix\url{https://www.cbc.ca/news/science/alt-tech-platforms-resurface-plandemic-1.5577013}.

\bibitem[{Bond(2022)}]{Bond2022}
Bond, S. 2022.
\newblock {Kicked off Facebook and Twitter, far-right groups lose online
  clout}.
\newblock
  \urlprefix\url{https://www.npr.org/2022/01/06/1070763913/kicked-off-facebook-and-twitter-far-right-groups-lose-online-clout}.

\bibitem[{Buckley and Schafer(2021)}]{Buckley2021}
Buckley, N.; and Schafer, J.~S. 2021.
\newblock {“Censorship-free” platforms: Evaluating content moderation
  policies and practices of alternative social media}.
\newblock \doi{10.31235/osf.io/yf9qz}.
\newblock \urlprefix\url{https://osf.io/preprints/socarxiv/yf9qz/}.

\bibitem[{Buntain et~al.(2021{\natexlab{a}})Buntain, Bonneau, Nagler, and
  Tucker}]{Buntain2021}
Buntain, C.; Bonneau, R.; Nagler, J.; and Tucker, J.~A. 2021{\natexlab{a}}.
\newblock {Youtube recommendations and effects on sharing across online social
  platforms}.
\newblock \emph{PACM on Human-Computer Interaction} 0(0).
\newblock ISSN 23318422.
\newblock \doi{10.1145/3449085}.
\newblock \urlprefix\url{https://doi.org/10.1145/3449085}.

\bibitem[{Buntain et~al.(2021{\natexlab{b}})Buntain, Buntain, Mccreadie, and
  Soboroff}]{Buntain2021a}
Buntain, C.; Buntain, C.; Mccreadie, R.; and Soboroff, I. 2021{\natexlab{b}}.
\newblock {Incident Streams 2020 : TRECIS in the Time of COVID-19}.
\newblock \emph{18th International Conference on Information Systems for Crisis
  Response and Management} (April).

\bibitem[{Buntain et~al.(2021{\natexlab{c}})Buntain, Innes, Mitts, and
  Shapiro}]{Buntain2021b}
Buntain, C.; Innes, M.; Mitts, T.; and Shapiro, J.~N. 2021{\natexlab{c}}.
\newblock {Effects of the Post-January 6 Deplatforming on Social Media
  Discourse}.

\bibitem[{Chu et~al.(2022)Chu, Arunasalam, Ozmen, and Celik}]{277236}
Chu, A.; Arunasalam, A.; Ozmen, M.~O.; and Celik, Z.~B. 2022.
\newblock {Behind the Tube: Exploitative Monetization of Content on {YouTube}}.
\newblock In \emph{31st USENIX Security Symposium (USENIX Security 22)}.
  Boston, MA: USENIX Association.
\newblock
  \urlprefix\url{https://www.usenix.org/conference/usenixsecurity22/presentation/chu}.

\bibitem[{David(2020)}]{David2020}
David, G. 2020.
\newblock {BitChute: Platforming Hate and Terror in the Uk}.
\newblock Technical report, HOPE not hate Charitable Trust.

\bibitem[{Davidson et~al.(2017)Davidson, Warmsley, Macy, and
  Weber}]{ICWSM1715665}
Davidson, T.; Warmsley, D.; Macy, M.; and Weber, I. 2017.
\newblock {Automated Hate Speech Detection and the Problem of Offensive
  Language}.
\newblock
  \urlprefix\url{https://aaai.org/ocs/index.php/ICWSM/ICWSM17/paper/view/15665/14843}.

\bibitem[{Doesburg(2021)}]{Doesburg2021}
Doesburg, J.~V. 2021.
\newblock \emph{{An alternative rabbit hole? An analysis of the construction of
  echo chambers within coronavirus activism groups on Telegram}}.
\newblock Masters, Utrecht University.

\bibitem[{Fair and Wesslen(2019)}]{Fair2019}
Fair, G.; and Wesslen, R. 2019.
\newblock {Shouting into the void: A database of the alternative social media
  platform Gab}.
\newblock \emph{Proceedings of the 13th International Conference on Web and
  Social Media, ICWSM 2019} (Icwsm): 608--610.

\bibitem[{Fink(2018)}]{10.2307/26508117}
Fink, C. 2018.
\newblock {Dangerous Speech, Anti-Muslim Violence, and Facebook In Myanmar}.
\newblock \emph{Journal of International Affairs} 71(1.5): 43--52.
\newblock ISSN 0022197X.
\newblock \urlprefix\url{https://www.jstor.org/stable/26508117}.

\bibitem[{Forberg(2021)}]{Forberg2021}
Forberg, P.~L. 2021.
\newblock {From the Fringe to the Fore: An Algorithmic Ethnography of the
  Far-Right Conspiracy Theory Group QAnon}.
\newblock \emph{Journal of Contemporary Ethnography} 1--27.
\newblock ISSN 15525414.
\newblock \doi{10.1177/08912416211040560}.

\bibitem[{Freelon, Marwick, and Kreiss(2020)}]{Freelon2020a}
Freelon, D.; Marwick, A.; and Kreiss, D. 2020.
\newblock {False equivalencies: Online activism from left to right}.
\newblock \emph{Science} 369(6508): 1197--1201.
\newblock ISSN 10959203.
\newblock \doi{10.1126/SCIENCE.ABB2428}.

\bibitem[{Heilweil(2021)}]{Heilweil2021}
Heilweil, R. 2021.
\newblock {Platforms are cracking down hard on political misinformation, but
  it's still easy to find}.
\newblock
  \urlprefix\url{https://www.vox.com/recode/22240408/facebook-youtube-twitter-qanon-misinformation-inauguration}.

\bibitem[{Hoseini et~al.(2021)Hoseini, Melo, Benevenuto, Feldmann, and
  Zannettou}]{Hoseini2021}
Hoseini, M.; Melo, P.; Benevenuto, F.; Feldmann, A.; and Zannettou, S. 2021.
\newblock {On the Globalization of the QAnon Conspiracy Theory Through
  Telegram}.
\newblock In \emph{arXiv preprint}.
\newblock \urlprefix\url{http://arxiv.org/abs/2105.13020}.

\bibitem[{J{\'{u}}nior et~al.(2021)J{\'{u}}nior, Melo, {Da Silva}, Benevenuto,
  and Almeida}]{Junior2021}
J{\'{u}}nior, M.; Melo, P.; {Da Silva}, A. P.~C.; Benevenuto, F.; and Almeida,
  J. 2021.
\newblock {Towards Understanding the Use of Telegram by Political Groups in
  Brazil}.
\newblock \emph{ACM International Conference Proceeding Series} 237--244.
\newblock \doi{10.1145/3470482.3479640}.

\bibitem[{Kearney, Chiang, and Massey(2020)}]{Kearney2020}
Kearney, M.~D.; Chiang, S.~C.; and Massey, P.~M. 2020.
\newblock {The Twitter origins and evolution of the COVID-19 “plandemic”
  conspiracy theory}.
\newblock \emph{Harvard Kennedy School Misinformation Review} 1(October):
  1--18.
\newblock \doi{10.37016/mr-2020-42}.

\bibitem[{Kopf(2020)}]{Kopf2020}
Kopf, S. 2020.
\newblock {“Rewarding Good Creators”: Corporate Social Media Discourse on
  Monetization Schemes for Content Creators}.
\newblock \emph{Social Media and Society} 6(4).
\newblock ISSN 20563051.
\newblock \doi{10.1177/2056305120969877}.

\bibitem[{Marshall(2020)}]{Marshall2020}
Marshall, J.~M. 2020.
\newblock {The Modern Memory Hole: Cyberethics Unchained}.
\newblock \emph{Athenaeum Review} 1(3).
\newblock
  \urlprefix\url{https://athenaeumreview.org/essay/the-modern-memory-hole-cyberethics-unchained/}.

\bibitem[{{Masters of Media}(2020)}]{MastersofMedia2020}
{Masters of Media}. 2020.
\newblock {Deplatforming and the rise of the alt-tech video hosting platform
  BitChute in 2020}.
\newblock Technical report, University of Amsterdam.
\newblock
  \urlprefix\url{http://mastersofmedia.hum.uva.nl/blog/2020/09/27/deplatforming-and-bitchute/}.

\bibitem[{M{\"{u}}ller and Schwarz(2021)}]{Muller2021}
M{\"{u}}ller, K.; and Schwarz, C. 2021.
\newblock {Fanning the Flames of Hate: Social Media and Hate Crime}.
\newblock \emph{Journal of the European Economic Association} 19(4):
  2131--2167.
\newblock ISSN 1542-4766.
\newblock \doi{10.1093/jeea/jvaa045}.

\bibitem[{Munger and Phillips(2019)}]{Munger_2019}
Munger, K.; and Phillips, J. 2019.
\newblock {A Supply and Demand Framework for YouTube Politics}.
\newblock In \emph{OSF Preprint}. OSF.
\newblock \urlprefix\url{osf.io/4wk63}.

\bibitem[{Olteanu et~al.(2014)Olteanu, Castillo, Diaz, and
  Vieweg}]{Olteanu2014}
Olteanu, A.; Castillo, C.; Diaz, F.; and Vieweg, S. 2014.
\newblock {CrisisLex: A Lexicon for Collecting and Filtering Microblogged
  Communications in Crises}.
\newblock \emph{Proc. of the 8th International Conference on Weblogs and Social
  Media} 376.
\newblock \doi{10.1.1.452.7691}.
\newblock
  \urlprefix\url{http://www.aaai.org/ocs/index.php/ICWSM/ICWSM14/paper/download/8091/8138}.

\bibitem[{Papadamou(2021)}]{Papadamou2021}
Papadamou, K. 2021.
\newblock \emph{{Characterizing Abhorrent, Misinformative, and Mistargeted
  Content on YouTube}}.
\newblock Doctoral thesis, Cyprus University of Technology.
\newblock \urlprefix\url{http://arxiv.org/abs/2105.09819}.

\bibitem[{Paudel et~al.(2021)Paudel, Blackburn, {De Cristofaro}, Zannettou, and
  Stringhini}]{Paudel2021}
Paudel, P.; Blackburn, J.; {De Cristofaro}, E.; Zannettou, S.; and Stringhini,
  G. 2021.
\newblock {An Early Look at the Gettr Social Network}
  \urlprefix\url{http://arxiv.org/abs/2108.05876}.

\bibitem[{Pelzer(2018)}]{Pelzer2018}
Pelzer, R. 2018.
\newblock {Policing of Terrorism Using Data from Social Media}.
\newblock \emph{European Journal for Security Research} 3(2): 163--179.
\newblock ISSN 2365-0931.
\newblock \doi{10.1007/s41125-018-0029-9}.
\newblock \urlprefix\url{https://doi.org/10.1007/s41125-018-0029-9}.

\bibitem[{Qazi, Imran, and Ofli(2020)}]{10.1145/3404820.3404823}
Qazi, U.; Imran, M.; and Ofli, F. 2020.
\newblock {GeoCoV19: A Dataset of Hundreds of Millions of Multilingual COVID-19
  Tweets with Location Information}.
\newblock \emph{SIGSPATIAL Special} 12(1): 6--15.
\newblock \doi{10.1145/3404820.3404823}.
\newblock \urlprefix\url{https://doi.org/10.1145/3404820.3404823}.

\bibitem[{Rauchfleisch and Kaiser(2021)}]{Rauchfleisch2021}
Rauchfleisch, A.; and Kaiser, J. 2021.
\newblock {Deplatforming the Far-right: An Analysis of YouTube and BitChute}.
\newblock \emph{SSRN Electronic Journal} (108).
\newblock \doi{10.2139/ssrn.3867818}.

\bibitem[{Rogers(2020)}]{Rogers2020}
Rogers, R. 2020.
\newblock {Deplatforming: Following extreme Internet celebrities to Telegram
  and alternative social media}.
\newblock \emph{European Journal of Communication} 35(3): 213--229.
\newblock ISSN 14603705.
\newblock \doi{10.1177/0267323120922066}.

\bibitem[{Rye, Blackburn, and Beverly(2020)}]{Rye2020}
Rye, E.; Blackburn, J.; and Beverly, R. 2020.
\newblock {Reading In-Between the Lines: An Analysis of Dissenter}.
\newblock \emph{Proceedings of the ACM SIGCOMM Internet Measurement Conference,
  IMC} 133--146.
\newblock \doi{10.1145/3419394.3423615}.

\bibitem[{Sharma(2021)}]{Sharma2021}
Sharma, A. 2021.
\newblock {Anonymous Leaked a Bunch of Data From a Right-Wing Web Host}.
\newblock
  \urlprefix\url{https://arstechnica.com/information-technology/2021/09/anonymous-leaks-gigabytes-of-data-from-epik-web-host-of-gab-and-parler/}.

\bibitem[{Thompson(2011)}]{10.2307/26463917}
Thompson, R. 2011.
\newblock {Radicalization and the Use of Social Media}.
\newblock \emph{Journal of Strategic Security} 4(4): 167--190.
\newblock ISSN 19440464, 19440472.
\newblock \urlprefix\url{http://www.jstor.org/stable/26463917}.

\bibitem[{Trujillo et~al.(2020)Trujillo, Gruppi, Buntain, and
  Horne}]{trujillo2020bitchute}
Trujillo, M.; Gruppi, M.; Buntain, C.; and Horne, B.~D. 2020.
\newblock {What is BitChute? Characterizing the "Free Speech" Alternative to
  YouTube}.
\newblock In \emph{Proceedings of the 31st ACM Conference on Hypertext and
  Social Media}, HT '20, 139--140. New York, NY, USA: Association for Computing
  Machinery.
\newblock ISBN 9781450370981.
\newblock \doi{10.1145/3372923.3404833}.
\newblock \urlprefix\url{https://doi.org/10.1145/3372923.3404833}.

\bibitem[{Tucker et~al.(2018)Tucker, Guess, Barber, Vaccari, Nyhan, Seigel,
  Sanovich, and Stukal}]{Tucker2018}
Tucker, J.; Guess, A.; Barber, P.; Vaccari, C.; Nyhan, B.; Seigel, A.;
  Sanovich, S.; and Stukal, D. 2018.
\newblock {Social Media, Political Polarization, and Political Disinformation:
  A Review of the Scientific Literature}.
\newblock Technical Report March, Hewlett Foundation.

\bibitem[{{Van Dijck}, de~Winkel, and Sch{\"{a}}fer(2021)}]{VanDijck2021}
{Van Dijck}, J.; de~Winkel, T.; and Sch{\"{a}}fer, M.~T. 2021.
\newblock {Deplatformization and the governance of the platform ecosystem}.
\newblock \emph{New Media and Society} (January).
\newblock ISSN 14617315.
\newblock \doi{10.1177/14614448211045662}.

\bibitem[{Warreth(2021)}]{Warreth2021}
Warreth, S. 2021.
\newblock \emph{{Comparing Far Right and Jihadi Use of Crowdfunding,
  Cryptocurrencies, and Blockchain Technology: Accessibility, Geography,
  Ideology}}.
\newblock Masters thesis, University of Glasgow.
\newblock \doi{10.13140/RG.2.2.27543.09123}.

\bibitem[{Wilson and Starbird(2021)}]{Wilson2021}
Wilson, T.; and Starbird, K. 2021.
\newblock {Cross-platform Information Operations: Mobilizing Narratives and
  Building Resilience through both 'Big' and 'Alt' Tech}.
\newblock \emph{Proceedings of the ACM on Human-Computer Interaction} 5(CSCW2).
\newblock ISSN 25730142.
\newblock \doi{10.1145/3476086}.

\bibitem[{Zannettou et~al.(2018)Zannettou, Bradlyn, {De Cristofaro}, Kwak,
  Sirivianos, Stringini, and Blackburn}]{zannettou2018gab}
Zannettou, S.; Bradlyn, B.; {De Cristofaro}, E.; Kwak, H.; Sirivianos, M.;
  Stringini, G.; and Blackburn, J. 2018.
\newblock {What is gab: A bastion of free speech or an alt-right echo chamber}.
\newblock In \emph{Companion Proceedings of the The Web Conference 2018},
  volume~12, 1007--1014.
\newblock ISBN 9781450356404.
\newblock ISSN 10367128.
\newblock \doi{10.1254/fpj.130.403}.

\bibitem[{Zignani et~al.(2019)Zignani, Quadri, Galdeman, Gaito, and
  Rossi}]{Zignani2019}
Zignani, M.; Quadri, C.; Galdeman, A.; Gaito, S.; and Rossi, G.~P. 2019.
\newblock {Mastodon content warnings: Inappropriate contents in a microblogging
  platform}.
\newblock \emph{Proceedings of the 13th International Conference on Web and
  Social Media, ICWSM 2019} (Icwsm): 639--645.

\end{thebibliography}
\end{small}

\appendix
\onecolumn
\section{Data Column Descriptions}\label{appendix:coumns}
In this appendix, we provide detailed descriptions of each data column in the \texttt{MeLa-BitChute} dataset. Below are tables for each table in the database (videos, comments, channels).

\begin{table*}[h]
\fontsize{9pt}{9pt}
\selectfont
\centering
\begin{tabular}{c|p{11cm}}
\textbf{Column Name} & \textbf{Description}\\
\toprule
url & URL to video\\\midrule
title & Title of the video\\\midrule
postdate & Date video was uploaded to BitChute. Unparsed from the website, like ``First published at 17:31 UTC on August 28th, 2020."\\\midrule
scrapedate & Date video was added to our database, Unix Epoch time, accuracy in seconds\\\midrule
description & Uploader-chosen description of their video\\\midrule
channel & URL to the channel\\\midrule
profile & URL to the uploader's profile. Note, a profile can have multiple channels, but a channel belongs to one profile. \\\midrule
category & Uploader picked category of the video. Note, the default category is `None' or `Other' depending on if the video comes before or after September 2020.\\\midrule
sensitivity & Uploader chosen sensititivity score, chosed from: ``Normal", ``NSFW", and ``NSFL"\\\midrule
views & Integer number of video views at data collection time. If the views data is $-1$, the video was removed by the uploader. If views is $-3$, the video was removed by BitChute. \\\midrule
view\_scrapedate & Unix Epoch time when views and comments were added to the database. Guaranteed to be at least one week after \texttt{scrapedate}.
\end{tabular}
\caption{\textbf{videos} data description}
\label{tbl:vids}
\end{table*}

\begin{table*}[h]
\fontsize{9pt}{9pt}
\selectfont
\centering
\begin{tabular}{c|p{11cm}}
\textbf{Column Name} & \textbf{Description}\\
\toprule
url & URL to video that the comment falls under\\\midrule
userid & A SHA256 hash that uniquely identifies each commenter \\\midrule
posthtml & The full HTML of the comment\\\midrule
posttext & The body text of the comment (a pre-processed version of \texttt{posthtml}) \\\midrule
comment\_id & A text ID identifying a comment on a video \\\midrule
parent\_id & If non-NULL, refers to the \texttt{comment\_id} of the parent comment
\end{tabular}
\caption{\textbf{comments} data description}
\label{tbl:cmts}
\end{table*}

\begin{table*}[h]
\fontsize{9pt}{9pt}
\selectfont
\centering
\begin{tabular}{c|p{11cm}}
\textbf{Column Name} & \textbf{Description}\\
\toprule
url & URL to the channel\\\midrule
scrapedate & UTC timestamp of the date on which the description data was collected\\\midrule
description & Full HTML of channel description on the `About' page of the channel. These range from very long descriptions to short or blank descriptions. If no description was found, the value of the column will be `Null'. However, note that occasionally we found channels with descriptions that were multiple blank characters, making the stored value not `Null'\\\midrule
description\_stripped & Text of description stripped from the HTML
\end{tabular}
\caption{\textbf{channels} data description}
\label{tbl:cmts_db}
\end{table*}

\end{document}